\documentclass[%
reprint,
showpacs,
 amsmath,amssymb,
aps,
prb,
]{revtex4-1}

\usepackage{graphicx}
\usepackage{dcolumn}
\usepackage{bm}

\begin{document}


\title{Homogenization of metasurfaces formed by random resonant particles\\ 
in periodical lattices}


\author{Andrei Andryieuski}
\email{andra@fotonik.dtu.dk}
\author{Andrei V. Lavrinenko}
\affiliation{%
 DTU Fotonik, Technical University of Denmark, \\ \O rsteds pl. 343, Kongens Lyngby DK-2800, Denmark
}%

\author{Mihail Petrov}%
\affiliation{%
 IRC Nanophotonics and Metamaterials, ITMO University, Birjevaja line V.O., 14, Saint-Petersburg 199034, Russia
}%

\author{Sergei A. Tretyakov}
\affiliation{
Department of Radio Science and Engineering, Aalto University,\\ PO Box 13000, Aalto FI-00076, Finland
}%

\date{\today}

\begin{abstract}
In this paper we suggest a simple analytical method for description of electromagnetic properties of a geometrically regular two-dimensional subwavelength arrays (metasurfaces) formed by particles with randomly fluctuating polarizabilities. Such metasurfaces are of topical  importance due to development of mass-scale bottom-up fabrication methods, for which fluctuations of the particles sizes, shapes, and/or composition are inevitable. Understanding and prediction of electromagnetic properties of such random metasurfaces  is a challenge. We propose an analytical homogenization method applicable for normal wave incidence on particles arrays with dominating electric dipole responses and validate it with numerical point-dipole modeling using the supercell approach. We demonstrate that fluctuations of particles polarizabilities lead to increased diffuse scattering despite the subwavelength lattice constant of the array. The proposed method can be readily extended to oblique incidence and particles with both electric and magnetic dipole resonances. 
\end{abstract}

\maketitle


\section{\label{sec:Intro} Introduction}

Exciting opportunities to control electromagnetic radiation with artificial composite materials with the properties on demand (metamaterials) \cite{Zheludev2010} have led to an explosion of ideas on the constituent elements designs as well as potential applications\cite{Zheludev2012}. In the most common case the building blocks (meta-atoms) are identical complex-shape resonant particles placed in a regular two-dimensional (metasurface) or three-dimensional (3D metamaterial) lattice. It became, however, obvious from the very beginning of the ``metamaterials era'' that a certain degree of randomness is inevitably present during fabrication. Moreover, cheap and mass-scale self-assembly methods\cite{Dintinger2012,Vignolini2011,Yang2014,electroless2015} are intensively developed for optical metamaterials as a replacement of more precise, but expensive deterministic nanolithography (electron beam lithography, focused ion beam milling, nanoimprint lithography). From this practical point of view there is a strong interest in the random systems not only in the area of metamaterials, but also broader \cite{Amorphous}, widening the topic of inhomogeneous composites, whose homogenization has been actively studied in the last one and a half centuries (see the overview in [\onlinecite{Sihvola1999}]).

Random metamaterials were investigated for applications in superlensing \cite{Nielsen2010}, electromagnetic absorbers \cite{Zhu2009,Andryieuski2014} and light-trapping structures \cite{Polman2015}. Optical properties of metamaterials with positional disorder were characterized experimentally and modeled numerically \cite{Helgert2009,Muhlig2011}. Subsequently, Albooyeh et al. suggested to account for additional scattering losses through the modified imaginary part of the dipoles interaction constant Im$(\beta)$ \cite{Albooyeh2012}. Later this approach was extended on a more general case of oblique incidence\cite{Albooyeh2014}. However, no any analytical connection between the correction coefficient to Im$(\beta)$ and disorder characteristics of the metamaterial was proposed.

In this work we propose an analytical method for calculating additional scattering losses of a metasurface consisting of random meta-atoms placed in a regular two-dimensional array. The analytical predictions are confirmed by numerical simulations employing the discrete dipole approximation with the supercell approach.

All dipole moments are assumed to be parallel to the external electric field vector, so we are using scalar representation instead of field and dipole moment vectors throughout the paper. We also use the optical convention for the exponential $\exp(-i\omega t)$.

\section{\label{sec:analyt} Analytical description of random metasurface}

\begin{figure*}
   \begin{center}
   \begin{tabular}{c}
   \centering
   \includegraphics[width=17cm]{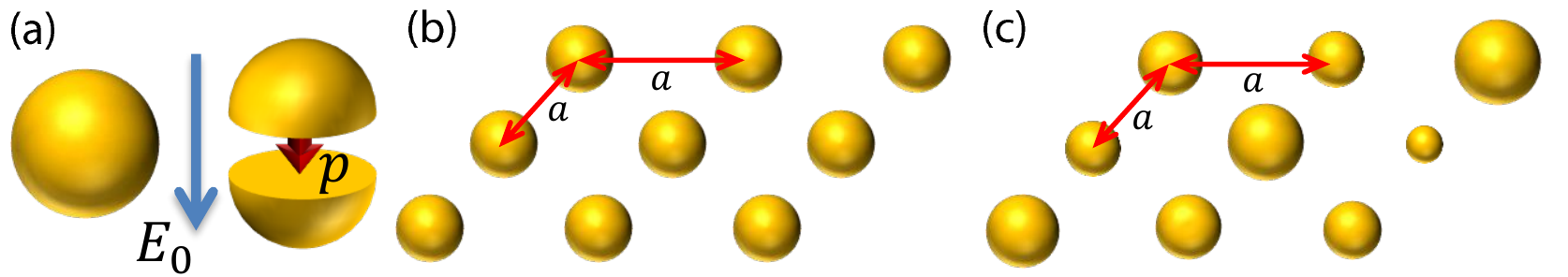}
   \end{tabular}
   \end{center}
   \caption{\label{Fig01} (Color online). (a) An incident plane wave with the amplitude $E_0$ excites dipole moment $p=\alpha_e E_0$ in an isolated meta-atom. (b) A regular array of similar particles. (c) A regular array of particles with random polarizabilities (for example, with random radii).}
\end{figure*}

We consider a regular two-dimensional array of meta-atoms (in other words, a metasurface). We assume that the considered particles behave as electric dipoles, possibly resonant, and the incident plane wave with amplitude $E_0$ excites only the electric dipole moment $p$, but neither a magnetic dipole nor higher-order multipoles [see Fig.~\ref{Fig01}(a)]. Using duality of the electric and magnetic fields the following methods and results can be easily generalized on the magnetic field and magnetic multipole moments. 

\subsection{Single particle}
\label{sec:single}

Conventionally a particle with an electric dipole is characterized by its polarizability $\alpha_e$, which is the proportionality coefficient between the dipole moment, induced in the particle, and the external electromagnetic field \cite{Tretyakov2003}. 
\begin{equation}
\label{eq:p}
	p=\alpha_e E_0.
\end{equation}
From the power balance equation (the power extracted from the "dipole moment source" is equal to the power radiated by the dipole) we get the well-known expression for Im$(\frac{1}{\alpha_e})$\cite{Tretyakov2003}:
\begin{equation}
\label{eq:DipoleExtractedPower}
	P_\mathrm{extr}=\frac{\omega}{2}\mathrm{Im}(p^* \cdot E)=-\frac{\omega}{2} \mathrm{Im}\left(\frac{1}{\alpha_e}\right) |p|^2,
\end{equation}
\begin{equation}
\label{eq:DipoleRadiatedPower}
    P_\mathrm{rad}=\frac{\omega^4|p|^2}{12\pi\varepsilon_0c^3},
\end{equation}
\begin{equation}
\label{eq:DipoleImInvAlpha}
    \mathrm{Im}\left(\frac{1}{\alpha_e}\right)=-\frac{k^3}{6\pi\varepsilon_0}.
\end{equation}
    
\noindent As we see, the power scattered by a dipole particle is determined by the imaginary part of the \emph{inverse} polarizability. That is why for the purposes of this study, where we will estimate the diffuse scattering from arrays of random-polarizability particles, it is  convenient to formulate the excitation problem in terms of the inverse polarizabilities. Thus, instead of operating with dipole moment $p$ excited by  electric field $E_0$ at the position of the dipole [see Fig. \ref{Fig01}(a)], we formulate the problem in a reverse way by addressing electric field $E$ excited at the dipole position by a virtual ``dipole moment source'' $p$. The excited electric field is
\begin{equation}
\label{eq:p_minus}
	E=-\frac{1}{\alpha_e} p,
\end{equation}

\noindent where the minus sign comes from the fact that by definition the dipole moment is directed from a negative charge to the positive charge, thus the excited electric field inside the particle is directed oppositely to the dipole moment.

\subsection{Regular 2D array of regular particles}

\label{sec:regular}

In case of a regular 2D square array [lattice constant $a<\lambda$, see Fig. \ref{Fig01}(b)], the dipole moment of a certain particle $p$ is related to the local electric field, which is the sum of the incident field and the interaction field induced by all the neighbors in the array $E_\mathrm{interaction}=\beta p$ (in regular arrays of identical particles dipole moment $p$ is the same for all particles), where $\beta$ is the so-called interaction constant, which can be estimated  analytically for $ka\leq1.5$ as

\begin{equation}
\label{eq:beta}
    \beta=\mathrm{Re}\left[\frac{ik}{4a^2\varepsilon_0}\left(1+\frac{1}{ikR_0}\right)e^{ikR_0}\right]+i\left(\frac{k}{2a^2\varepsilon_0}-\frac{k^3}{6\pi\varepsilon_0}\right),
\end{equation}

\noindent with $R_0=a/1.438$ \cite{Tretyakov2003}.

Considering the virtual "dipole moment sources" that support constant dipole moment amplitude $p$ in all particles, the electric field at the position of a certain dipole is
\begin{equation}
\label{eq:ArrayEfield}
	E=-\frac{1}{\alpha_e} p+\beta p,
\end{equation}

From the power per unit cell balance equation, we get
\begin{equation}
\label{eq:ArrayExtractedPower}
	P_\mathrm{extr}=\frac{\omega}{2}\mathrm{Im}(p^*\cdot E)=-\frac{\omega}{2} \mathrm{Im}\left(\frac{1}{\alpha_e}-\beta\right)|p|^2,
\end{equation}
\begin{equation}
\label{eq:ArrayRadiatedPower}
    P_\mathrm{rad}=\frac{\eta\omega^2|p|^2}{4a^2},
\end{equation}
\begin{equation}
\label{eq:ArrayImInvAlpha}
    \mathrm{Im}\left(\frac{1}{\alpha_e}-\beta\right)=-\frac{k}{2\varepsilon_0a^2},
\end{equation}

\noindent where $\eta$ is the free-space impedance, and we have taken into account that radiation from the subwavelength array may occur only into plane waves propagating to the opposite sides from the array normally to its plane. From here it is easy to get the exact expression for Im$(\beta)$:
\begin{equation}
\label{eq:ArrayImInvBeta}
    \mathrm{Im}(\beta)=\frac{k}{2\varepsilon_0a^2}-\frac{k^3}{6\pi\varepsilon_0}.
\end{equation}
If we introduce the normalized polarizability $\alpha_n=\alpha_e/\varepsilon_0a^3$ and the normalized interaction constant $\beta_n=\beta\varepsilon a^3$, then the amplitude reflection and transmission coefficients can be expressed as
\begin{eqnarray}
\label{eq:ArrayReflectionTransmission}
    r=\frac{ika}{2}\left(\frac{1}{\alpha_n}-\beta\right),\qquad
    t=1+r,
\end{eqnarray}

\noindent and the loss factor (in the case of subwavelength regular arrays only absorbance contributes to it) reads
\begin{eqnarray}
\label{eq:ArrayLosses}
    A=1-|r|^2-|t|^2.
\end{eqnarray}

\subsection{Regular 2D array of random particles}

\label{sec:random}

In case of a regular square array with the same lattice constant $a$, but with fluctuating polarizabilities $\alpha_\mathrm{fluct}$ of the particles [see Fig. \ref{Fig01}(c)] we again assume a regular array of virtual parallel ``dipole moment sources'' within each particle that create parallel random electric fields at the position of each inclusion, instead of considering a uniform incident electric field which creates parallel random dipole moments in the meta-atoms. The fluctuating electric field in the position of a certain particle is
\begin{equation}
\label{eq:RandomEfield}
	E=-\frac{1}{\alpha_\mathrm{fluct}}p+\beta p.
\end{equation}

\noindent We assume the randomness of the polarizabilities to be small ($|\Delta \frac{1}{\alpha}|\ll|\langle\frac{1}{\alpha}\rangle|$), what allows us to split the electric field into the average and fluctuating parts
\begin{equation}
\label{eq:RandomEfieldSplit}
\begin{split}
	E=\left(-\langle\frac{1}{\alpha}\rangle+\beta \right)p+\left(\langle\frac{1}{\alpha}\rangle-\frac{1}{\alpha_\mathrm{fluct}} \right)p = \\ = \left(-\langle\frac{1}{\alpha}\rangle+\beta \right)p-\Delta\frac{1}{\alpha}p.
\end{split}
\end{equation}
The average extracted power per unit cell is
\begin{equation}
\label{eq:RandomExtractedPower}
	P_\mathrm{extr}=\langle\frac{\omega}{2}\mathrm{Im}(p^*E)\rangle=-\frac{\omega}{2} \mathrm{Im}(\langle\frac{1}{\alpha}\rangle-\beta) |p|^2.
\end{equation}

The average part of the electric field in (\ref{eq:RandomEfieldSplit}) corresponds to radiation of plane waves as in the case of non-fluctuating particles, so the radiated power per unit cell is
\begin{equation}
\label{eq:RandomRadiatedPower}
    P_\mathrm{rad}=\frac{\eta\omega^2|p|^2}{4a^2},
\end{equation}

\noindent while the fluctuating part in (\ref{eq:RandomEfieldSplit}) should result in diffuse scattering. In order to estimate the power diffusely scattered by the array we replace the fluctuation of the electric field at the position of the dipoles with the fluctuation of the dipole moment of independent identical particles with the average inverse polarizability $\langle\frac{1}{\alpha}\rangle$
\begin{equation}
\label{eq:FluctuatingReplace}
    \Delta E=-p\left(\frac{1}{\alpha_\mathrm{fluct}}-\langle\frac{1}{\alpha}\rangle\right)=-\Delta p\langle\frac{1}{\alpha}\rangle,
\end{equation}
\begin{equation}
\label{eq:FluctuatingDipole}
   \Delta p= p\left(\frac{1}{\alpha_\mathrm{fluct}}-\langle\frac{1}{\alpha}\rangle\right)/\langle\frac{1}{\alpha}\rangle.
\end{equation}

\noindent Then the average power per unit cell radiated by the set of statistically independent dipole fluctuations reads
\begin{equation}
\label{eq:DipoleRadiatedPower2}
    P_\mathrm{diffuse}=\frac{\omega^4\langle|\Delta p|^2\rangle}{12\pi\varepsilon_0c^3},
\end{equation}

\noindent Using the power balance equation we finally get
\begin{equation}
\label{eq:RandomImInvBeta}
    \mathrm{Im}(\beta)=\frac{k}{2\varepsilon_0a^2}-\frac{k^3}{6\pi\varepsilon_0}\left(1-\left\langle\left|\frac{\dfrac{1}{\alpha_\mathrm{fluct}}}{\langle\dfrac{1}{\alpha}\rangle}-1\right|^2\right\rangle\right).
\end{equation}

\noindent It is convenient to introduce a factor 
\begin{equation}
\label{eq:RandomFactor}
    r_n=1-\left\langle\left|\frac{\dfrac{1}{\alpha_\mathrm{fluct}}}{\langle\dfrac{1}{\alpha}\rangle}-1\right|^2\right\rangle=1-\Delta,
\end{equation}
\noindent where $\Delta=\left\langle\left|\frac{{1}/{\alpha_\mathrm{fluct}}}{\langle{1}/{\alpha}\rangle}-1\right|^2\right\rangle$. Factor $r_n$ is analogous to the randomness factor introduced in [\onlinecite{Albooyeh2012}] for position-disordered metamaterials, since it relates the diffuse scattering losses to the interaction constant $\beta$. The loss factor calculated with formulas (\ref{eq:ArrayReflectionTransmission})-(\ref{eq:ArrayLosses}), but with the modified $\beta$ according to  formula (\ref{eq:RandomImInvBeta}), gives the total losses (absorption and diffuse scattering)
\begin{equation}
\label{eq:RandomLossFactor}
    A=1-|r|^2-|t|^2.
\end{equation}

In case of lossless spherical isotropic particles  of radius $R$ at the frequencies far from the plasmonic resonance, where $(\varepsilon+2)/(\varepsilon-1) \approx 1$, the normalized inverse polarizability, interaction constant and their difference read 
\begin{equation}
\label{eq:InversePolarizLosslessSphereMetal}
    \frac{1}{\alpha_n}=\textrm{Re}(\frac{1}{\alpha_n})-i\frac{k^3a^3}{6\pi},
\end{equation}
\begin{equation}
\label{eq:BetaLosslessSphereMetal}
    \beta_n=\textrm{Re}(\beta_n)+i\frac{ka}{2}-i\frac{k^3a^3}{6\pi}(1-\Delta).
\end{equation}
\begin{equation}
\label{eq:InversePolarizBetaLosslessSphereMetal}
\begin{split}
    \frac{1}{\alpha_n}-\beta_n=\textrm{Re}(\frac{1}{\alpha_n})-\textrm{Re}(\beta_n)-i\frac{ka}{2}-i\frac{k^3a^3}{6\pi}\Delta \\
    =X-i\left(\frac{ka}{2}-\frac{k^3a^3}{6\pi}\Delta\right),
\end{split}
\end{equation}
\noindent where $X=\textrm{Re}(\frac{1}{\alpha_n}-\beta_n)$. For $X\gg ka$, i.e. the common case far from the resonance and small randomness $\Delta\ll1$

\begin{equation}
\label{eq:LossFactorLosslessSphereMetal}
    A\approx \frac{k^4a^4\Delta}{6\pi X^2},
\end{equation}

\begin{figure*}
   \begin{center}
   \begin{tabular}{c}
   \includegraphics[width=17cm]{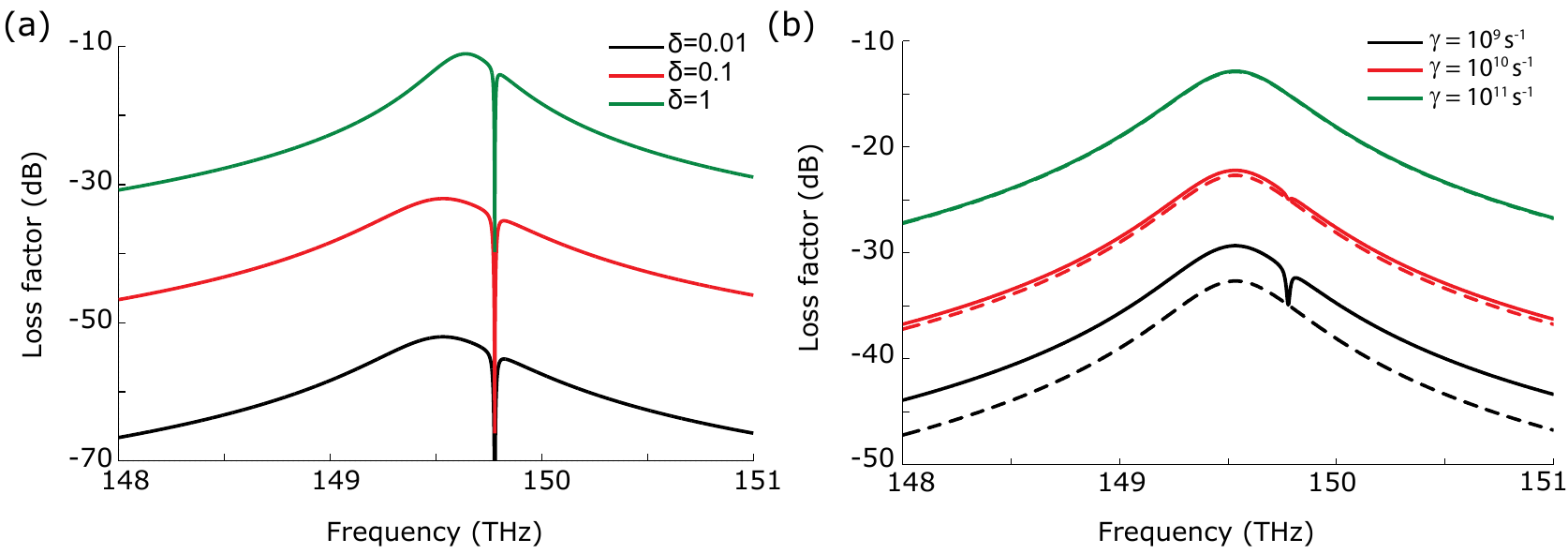}
   \end{tabular}
   \end{center}
   \caption{\label{Fig02} (Color online). (a) Diffuse scattering losses for a lossless particles array with the randomness $\delta =$ 0.01 (black), 0.1 (red) and 1 (green). (b) Total loss factor for the array of random (solid) and regular (dashed) lossy particles with the material damping $\gamma = 10^9$ s$^{-1}$ (black), $10^{10}$ s$^{-1}$ (red) and $10^{11}$ s$^{-1}$ (green).}
\end{figure*} 

\noindent which means that the loss factor (diffuse scattering in this case, since the particles are lossless) is proportional to the randomness factor $\Delta$ and inversely proportional to $\lambda^4$ in agreement with the assumption of the Rayleigh scattering regime.

As an illustrative example we consider an array of spherical particles of radius $R$ homogeneously distributed in the interval $[R_\textrm{mean}(1-\delta/2);R_\textrm{mean}(1+\delta/2)]$ ($\delta$ is the ``size randomness'' parameter) around the mean value of the radius $R_\textrm{mean}=20$ nm and arranged in a regular square array with lattice constant $a=200$ nm. The particles are made of a plasmonic Drude material with plasma frequency $\omega_P=1.63\times10^{15}$ s$^{-1}$ and suspended in free space. The considered frequency range of interest lies around the plasmonic resonance frequency for an isolated spherical particle $f_0=\omega_P/2\pi\sqrt{3}=150$ THz corresponding to the wavelength $\lambda = 2$ $\mu$m. The array is well subwavelength ($a\approx \lambda/10$).

In case of lossless particles, there are no losses (neither Ohmic, nor diffuse scattering) in the regular array of non-fluctuating particles. With increasing randomness [see Fig. \ref{Fig02}(a)], the diffuse scattering losses grow up. The curves experience the maximum close to the resonant frequency, which is quite natural, since the amplitude of the excited dipole moment as well as the fluctuations are maximal around the resonance. At frequency 149.8 THz a sharp dip in scattering losses is observed. This dip happens exactly on the plasmonic resonance frequency ($\mathrm{Re}(1/\alpha_e)=0$) and is related to the specific type of randomization, that is particles radius randomness. Indeed, the inverse polarizability of a spherical particle in the quasi-static limit is
\begin{equation}
\label{eq:InvAlpha}
    \frac{1}{\alpha_e}=\frac{1}{4\pi\varepsilon_0R^3}\frac{\varepsilon+2}{\varepsilon-1}-i\frac{k^3}{6\pi\varepsilon_0}.
\end{equation}

In case of absense of dissipation losses and at the resonance ($\mathrm{Re}(1/\alpha_e)=0$) the inverse polarizabilities of all particles become identical and equal to $\frac{1}{\alpha_e}=-i\frac{k^3}{6\pi\varepsilon_0}$, thus, no diffuse scattering can occur at this frequency, and the loss factor $A = 0$.

In case of lossy particles [see Fig.~\ref{Fig02}(b)] the total losses calculated with the formulas (\ref{eq:ArrayReflectionTransmission})-(\ref{eq:ArrayLosses}) include both Ohmic and scattering losses. Comparing these total losses [Fig. \ref{Fig02}(b), solid lines] with the Ohmic losses for the array of regular particles (dashed lines), we see that diffuse scattering losses dominate in case of small material losses below $\gamma=10^{10}$ s$^{-1}$. For comparison, material losses of noble metals are in the order of $\gamma\approx 10^{14}-10^{15}$ s$^{-1}$ and of graphene of $\gamma\approx 10^{12}-10^{14}$ s$^{-1}$.

\section{\label{sec:numer} Numerical supercell approach}

\begin{figure*}
   \begin{center}
   \begin{tabular}{c}
   \includegraphics[width=17cm]{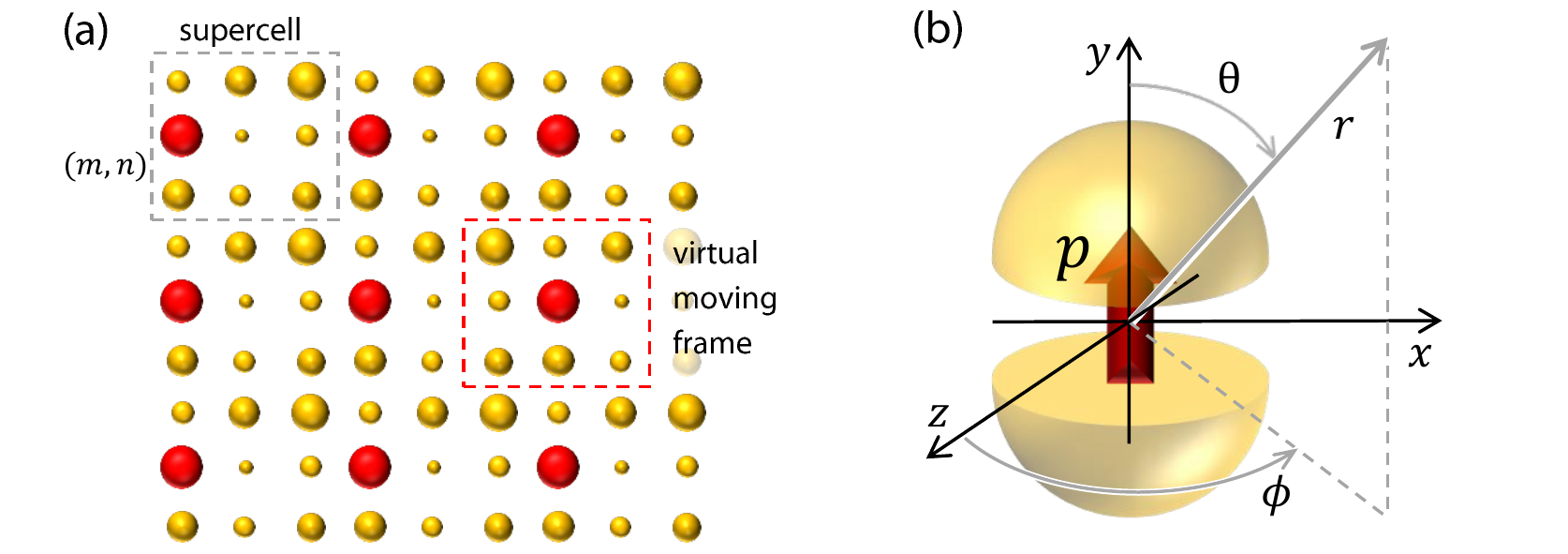}
   \end{tabular}
   \end{center}
   \caption{\label{Fig03} (Color online). (a) Illustration of the supercell modeling approach: particles are random within each supercell; the supercells are periodically repeated. (b) Coordinate system for calculating the power scattered by a dipole.}
\end{figure*}

In order to test the proposed analytical model of diffuse scattering losses, we performed  modeling of a regular array of random particles using the supercell approach. Since the effects are clearly observable for small Ohmic losses, we considered the lossless particles case only. In the frame of the supercell approach the particles are random within each supercell, and the supercells are periodically repeated in space [see the illustration in Fig. \ref{Fig03}(a)]. 

Since the respective dipoles in various supercells are identical [for example, the red circles on Fig. \ref{Fig03}(a)], this allows us to take into account all the interactions within the infinite quasi-random array as well as with the incident plane wave. Thus, we are able to determine the dipole moments of all $N\times N$ particles within a supercell by solving a  system of $N^2$ linear equations (refer to Appendix 1 for details). 

After knowing all the dipole moments we are able to calculate the scattered fields of each dipole [see Fig. \ref{Fig03}(b)] as well as the total intensity of transmitted, reflected or diffusely scattered electromagnetic waves by the complete infinite array (refer to Appendix 2 for details). In order to properly simulate randomness, the supercells are larger than the wavelength and being periodically repeated in space the supercells form a two-dimensional diffraction grating, so that only specific directions for scattering are allowed. 

\begin{figure}
   \begin{center}
   \begin{tabular}{c}
   \includegraphics[width=8cm]{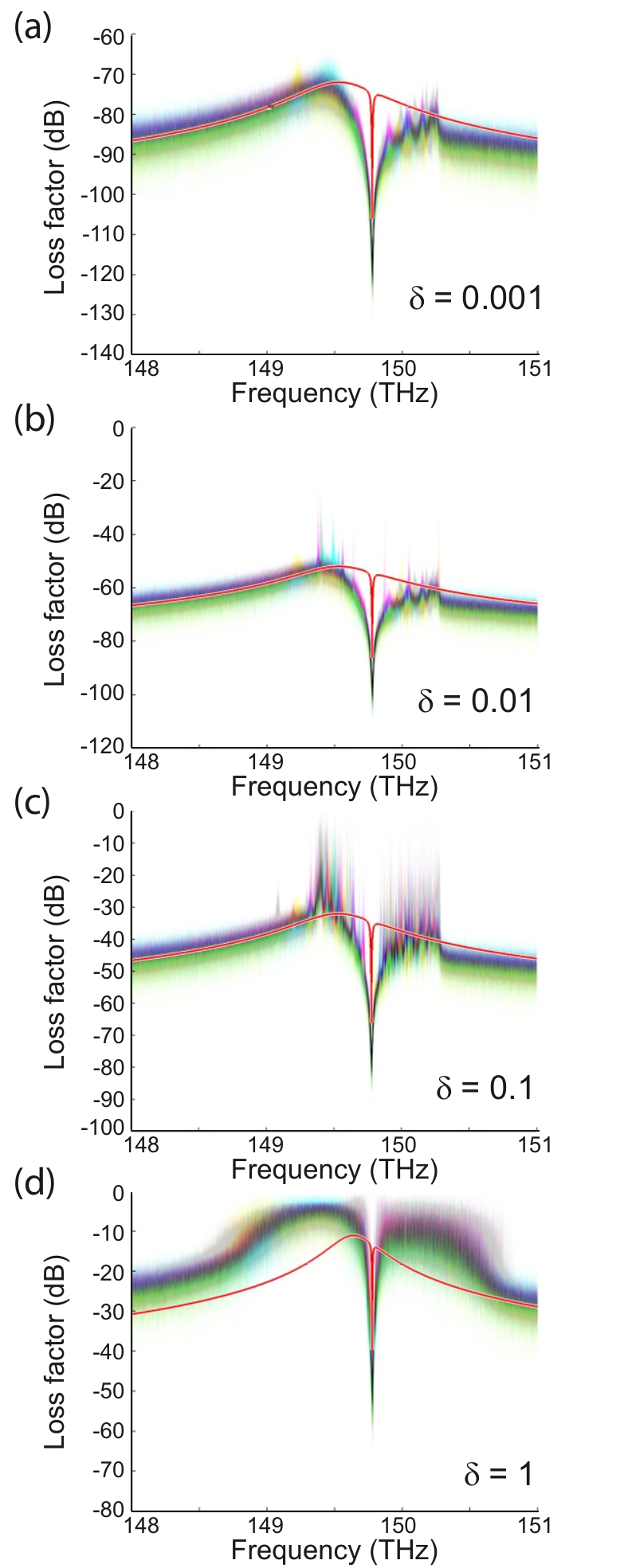}
   \end{tabular}
   \end{center}
   \caption{\label{Fig04} (Color online). Loss factor for the  regular array of random lossless particles with the randomness $\delta = 0.001$ (a), $\delta = 0.01$ (b), $\delta = 0.1$ (c) and $\delta = 1$ (d). Results of numerical simulations (high transparency lines) for 500 realizations of the supercells of size 13x13 (cyan), 15x15 (yellow), 17x17 (magenta) and 19x19 (grey) are compared to the analytical prediction (red line).}
\end{figure}

\begin{figure}
   \begin{center}
   \begin{tabular}{c}
   \includegraphics[width=8cm]{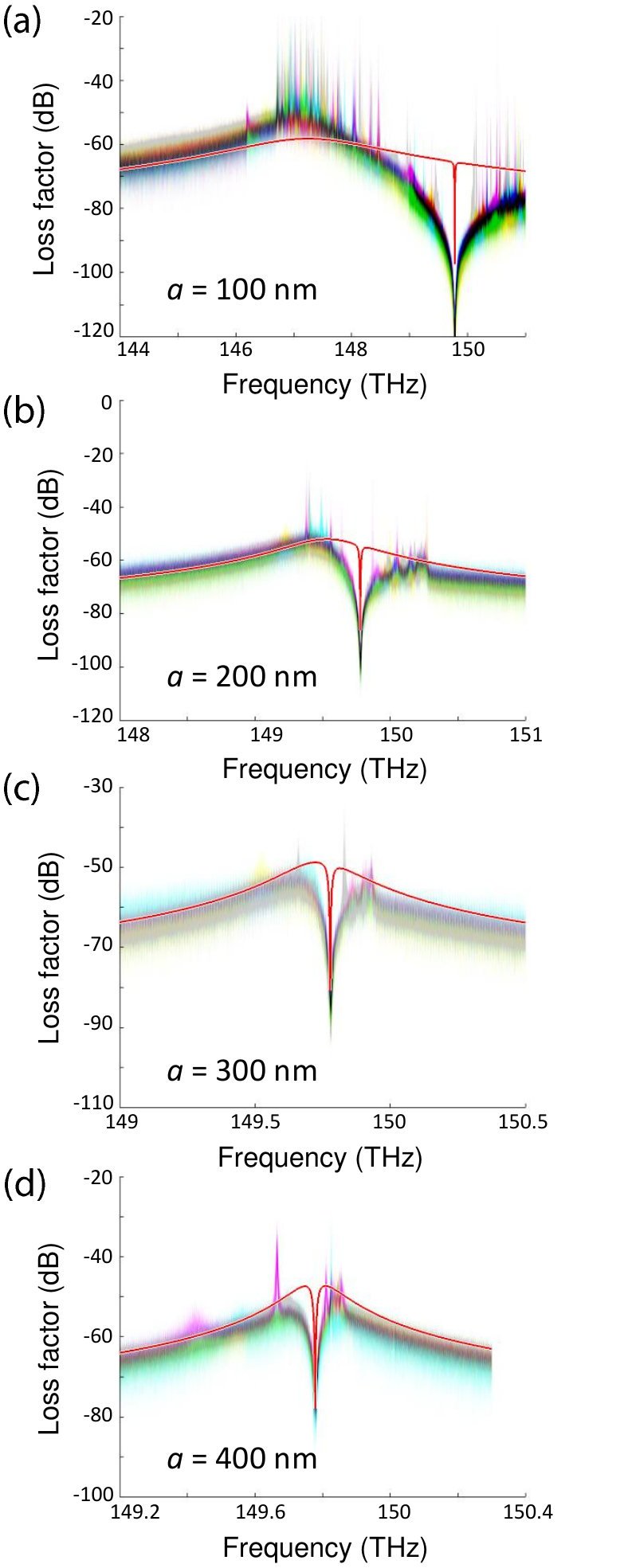}
   \end{tabular}
   \end{center}
   \caption{\label{Fig05} (Color online). Loss factor for the  regular array of random lossless particles with the lattice constant $a = 100$ nm (a), $a = 200$ nm (b), $a = 300$ nm (c) and $a = 400$ nm (d). Results of numerical simulations (high transparency lines) for 500 realizations of the supercells of size 13x13 (cyan), 15x15 (yellow), 17x17 (magenta) and 19x19 (grey) are compared to the analytical prediction (red line).}
\end{figure}

We used supercells with the sizes from $13\times 13$ to $19\times 19$ particles. The results of 2000 random realizations (500 realizations of the each supercells of sizes $13\times13$, $15\times15$, $17\times17$ and $19\times19$) are overlapped in Figs. \ref{Fig04} and \ref{Fig05} and compared to the analytical prediction. Even though such quasi-random supercells arrays are a kind of two-dimensional diffraction gratings, and the diffuse scattering is allowed to certain directions only, consideration of various supercells sizes gives a good approximation for a truly random infinite array.

The results for different supercell sizes are not identical, as can be seen from the coloring of specific parts of the graphs in Figs.~\ref{Fig04}-\ref{Fig05}. Namely, cyan corresponds to 13x13, yellow to 15x15, magenta to 17x17 and grey to 19x19 supercell size, while green, red, and blue colors are the results of overlapping cyan and yellow, yellow and magenta, and magenta and cyan, correspondingly. The darkest area represents the most probable value of the loss factor.

We observe not only qualitative, but also quantitative agreement between the numerical and analytical curves, reproducing the maximum of diffuse scattering, except for the case of large size randomness factor $\delta=1$ [Fig.~\ref{Fig04}(d)], for which the analytical approach underestimates the diffuse scattering losses. The resonant dip corresponding to the plasmonic resonance is observed at the same frequencies for analytical and numerical curves in all cases. The dip width in numerical modeling is, however, larger than the analytical formulas predict and that requires further investigation. 

The agreement between analytics and numerics is nearly perfect away from the resonance. However, in the very proximity to the plasmonic resonance the numerical spectra exhibit sharp features that are more pronounced for larger randomness and are most probably related to the effects of localized resonances excitation in a two-dimensional array.

Overall, the values of the loss factor due to diffuse scattering are small for small randomness [for example, -70 dB for $\delta = 0.001$, see Fig.~\ref{Fig04} (a)] but can be quite significant for larger randomness [see Figs. 4 (c) and (d)].

\section{\label{sec:discuss} Discussion and conclusions}

In summary, we introduced a simple analytical model  of diffuse scattering in periodical  two-dimensional arrays of particles with random polarizabilities in each unit cell. The diffuse scattering is estimated through a correction of the imaginary part of the interaction constant Im($\beta$) and is related to the statistics of the random particles polarizabilities. We considered fluctuations of particle radius, but the derived formulas are general and can be applied to any possible fluctuations of particles geometrical and material parameters. The analytical formula predicts a resonant increase of the diffuse scattering losses in the vicinity of the plasmonic resonance and this effect is confirmed by  numerical modeling with the supercell approach. 

We would like to emphasize the fact that the array under consideration is deeply subwavelength ($a\approx \lambda/20-\lambda/5$ in Fig. \ref{Fig05}) and for subwavelength regular arrays of identical scatterers diffuse scattering cannot occur in principle. The observed effects can be explained, simplistically, by stating that even a deeply-subwavelength regular array of particles with random polarizabilities becomes effectively inhomogeneous also at the wavelength scale, making diffuse scattering possible.

The proposed analytical approach underestimates diffuse scattering losses for large randomness and this is expected, since the analytical description was formulated from the very beginning for small deviations from a regular array of identical particles. For very large randomness other methods should be used.

In most practical situations, the material loss contribution to the total loss factor are typically much larger than the diffuse scattering due to randomness of the polarizabilities. The diffuse scattering can, however, be considerable for arrays of low-loss all-dielectric metamaterials, which are currently attracting considerable attention of the scientific community. We believe that the suggested simple approach will be especially useful for the description of all-dielectric random metamaterials and metasurfaces.

\section*{Acknowledgments}
A.A. acknowledges  support from the Danish Council for Independent Research via the GraTer project (Contract No. 0602-02135B). S.A.T. acknowledges support for his sabbatical leave from Otto M\o nsteds Fond, Technical University of Denmark, and Aalto University.

\bibliographystyle{apsrev4-1}
\bibliography{bibliography}

\begin{thebibliography}{17}%
\makeatletter
\providecommand \@ifxundefined [1]{%
 \@ifx{#1\undefined}
}%
\providecommand \@ifnum [1]{%
 \ifnum #1\expandafter \@firstoftwo
 \else \expandafter \@secondoftwo
 \fi
}%
\providecommand \@ifx [1]{%
 \ifx #1\expandafter \@firstoftwo
 \else \expandafter \@secondoftwo
 \fi
}%
\providecommand \natexlab [1]{#1}%
\providecommand \enquote  [1]{``#1''}%
\providecommand \bibnamefont  [1]{#1}%
\providecommand \bibfnamefont [1]{#1}%
\providecommand \citenamefont [1]{#1}%
\providecommand \href@noop [0]{\@secondoftwo}%
\providecommand \href [0]{\begingroup \@sanitize@url \@href}%
\providecommand \@href[1]{\@@startlink{#1}\@@href}%
\providecommand \@@href[1]{\endgroup#1\@@endlink}%
\providecommand \@sanitize@url [0]{\catcode `\\12\catcode `\$12\catcode
  `\&12\catcode `\#12\catcode `\^12\catcode `\_12\catcode `\%12\relax}%
\providecommand \@@startlink[1]{}%
\providecommand \@@endlink[0]{}%
\providecommand \url  [0]{\begingroup\@sanitize@url \@url }%
\providecommand \@url [1]{\endgroup\@href {#1}{\urlprefix }}%
\providecommand \urlprefix  [0]{URL }%
\providecommand \Eprint [0]{\href }%
\providecommand \doibase [0]{http://dx.doi.org/}%
\providecommand \selectlanguage [0]{\@gobble}%
\providecommand \bibinfo  [0]{\@secondoftwo}%
\providecommand \bibfield  [0]{\@secondoftwo}%
\providecommand \translation [1]{[#1]}%
\providecommand \BibitemOpen [0]{}%
\providecommand \bibitemStop [0]{}%
\providecommand \bibitemNoStop [0]{.\EOS\space}%
\providecommand \EOS [0]{\spacefactor3000\relax}%
\providecommand \BibitemShut  [1]{\csname bibitem#1\endcsname}%
\let\auto@bib@innerbib\@empty
\bibitem [{\citenamefont {Zheludev}(2010)}]{Zheludev2010}%
  \BibitemOpen
  \bibfield  {author} {\bibinfo {author} {\bibfnamefont {N.~I.}\ \bibnamefont
  {Zheludev}},\ }\href {\doibase 10.1126/science.1186756} {\bibfield  {journal}
  {\bibinfo  {journal} {Science}\ }\textbf {\bibinfo {volume} {328}},\ \bibinfo
  {pages} {582} (\bibinfo {year} {2010})},\ \Eprint
  {http://arxiv.org/abs/http://www.sciencemag.org/content/328/5978/582.full.pdf}
  {http://www.sciencemag.org/content/328/5978/582.full.pdf} \BibitemShut
  {NoStop}%
\bibitem [{\citenamefont {Zheludev}\ and\ \citenamefont
  {Kivshar}(2012)}]{Zheludev2012}%
  \BibitemOpen
  \bibfield  {author} {\bibinfo {author} {\bibfnamefont {N.~I.}\ \bibnamefont
  {Zheludev}}\ and\ \bibinfo {author} {\bibfnamefont {Y.~S.}\ \bibnamefont
  {Kivshar}},\ }\href@noop {} {\bibfield  {journal} {\bibinfo  {journal}
  {Nature materials}\ }\textbf {\bibinfo {volume} {11}},\ \bibinfo {pages}
  {917} (\bibinfo {year} {2012})}\BibitemShut {NoStop}%
\bibitem [{\citenamefont {Dintinger}\ \emph {et~al.}(2012)\citenamefont
  {Dintinger}, \citenamefont {M\"{u}hlig}, \citenamefont {Rockstuhl},\ and\
  \citenamefont {Scharf}}]{Dintinger2012}%
  \BibitemOpen
  \bibfield  {author} {\bibinfo {author} {\bibfnamefont {J.}~\bibnamefont
  {Dintinger}}, \bibinfo {author} {\bibfnamefont {S.}~\bibnamefont
  {M\"{u}hlig}}, \bibinfo {author} {\bibfnamefont {C.}~\bibnamefont
  {Rockstuhl}}, \ and\ \bibinfo {author} {\bibfnamefont {T.}~\bibnamefont
  {Scharf}},\ }\href@noop {} {\bibfield  {journal} {\bibinfo  {journal} {Opt.
  Mater. Express}\ }\textbf {\bibinfo {volume} {2}},\ \bibinfo {pages} {269}
  (\bibinfo {year} {2012})}\BibitemShut {NoStop}%
\bibitem [{\citenamefont {Vignolini}\ \emph {et~al.}(2012)\citenamefont
  {Vignolini}, \citenamefont {Yufa}, \citenamefont {Cunha}, \citenamefont
  {Guldin}, \citenamefont {Rushkin}, \citenamefont {Stefik}, \citenamefont
  {Hur}, \citenamefont {Wiesner}, \citenamefont {Baumberg},\ and\ \citenamefont
  {Steiner}}]{Vignolini2011}%
  \BibitemOpen
  \bibfield  {author} {\bibinfo {author} {\bibfnamefont {S.}~\bibnamefont
  {Vignolini}}, \bibinfo {author} {\bibfnamefont {N.~A.}\ \bibnamefont {Yufa}},
  \bibinfo {author} {\bibfnamefont {P.~S.}\ \bibnamefont {Cunha}}, \bibinfo
  {author} {\bibfnamefont {S.}~\bibnamefont {Guldin}}, \bibinfo {author}
  {\bibfnamefont {I.}~\bibnamefont {Rushkin}}, \bibinfo {author} {\bibfnamefont
  {M.}~\bibnamefont {Stefik}}, \bibinfo {author} {\bibfnamefont
  {K.}~\bibnamefont {Hur}}, \bibinfo {author} {\bibfnamefont {U.}~\bibnamefont
  {Wiesner}}, \bibinfo {author} {\bibfnamefont {J.~J.}\ \bibnamefont
  {Baumberg}}, \ and\ \bibinfo {author} {\bibfnamefont {U.}~\bibnamefont
  {Steiner}},\ }\href {\doibase 10.1002/adma.201103610} {\bibfield  {journal}
  {\bibinfo  {journal} {Advanced Materials}\ }\textbf {\bibinfo {volume}
  {24}},\ \bibinfo {pages} {OP23} (\bibinfo {year} {2012})}\BibitemShut
  {NoStop}%
\bibitem [{\citenamefont {Yang}\ \emph {et~al.}(2014)\citenamefont {Yang},
  \citenamefont {Ni}, \citenamefont {Yin}, \citenamefont {Kante}, \citenamefont
  {Zhang}, \citenamefont {Zhu}, \citenamefont {Wang},\ and\ \citenamefont
  {Zhang}}]{Yang2014}%
  \BibitemOpen
  \bibfield  {author} {\bibinfo {author} {\bibfnamefont {S.}~\bibnamefont
  {Yang}}, \bibinfo {author} {\bibfnamefont {X.}~\bibnamefont {Ni}}, \bibinfo
  {author} {\bibfnamefont {X.}~\bibnamefont {Yin}}, \bibinfo {author}
  {\bibfnamefont {B.}~\bibnamefont {Kante}}, \bibinfo {author} {\bibfnamefont
  {P.}~\bibnamefont {Zhang}}, \bibinfo {author} {\bibfnamefont
  {J.}~\bibnamefont {Zhu}}, \bibinfo {author} {\bibfnamefont {Y.}~\bibnamefont
  {Wang}}, \ and\ \bibinfo {author} {\bibfnamefont {X.}~\bibnamefont {Zhang}},\
  }\href@noop {} {\bibfield  {journal} {\bibinfo  {journal} {Nature
  nanotechnology}\ }\textbf {\bibinfo {volume} {9}},\ \bibinfo {pages} {1002}
  (\bibinfo {year} {2014})}\BibitemShut {NoStop}%
\bibitem [{\citenamefont {Durhuus}\ \emph {et~al.}(2015)\citenamefont
  {Durhuus}, \citenamefont {V.}, \citenamefont {Andryieuski}, \citenamefont
  {Pizzocchero}, \citenamefont {Bøggild}, \citenamefont {Malureanu},\ and\
  \citenamefont {Lavrinenko}}]{electroless2015}%
  \BibitemOpen
  \bibfield  {author} {\bibinfo {author} {\bibfnamefont {D.}~\bibnamefont
  {Durhuus}}, \bibinfo {author} {\bibfnamefont {L.~M.}\ \bibnamefont {V.}},
  \bibinfo {author} {\bibfnamefont {A.}~\bibnamefont {Andryieuski}}, \bibinfo
  {author} {\bibfnamefont {F.}~\bibnamefont {Pizzocchero}}, \bibinfo {author}
  {\bibfnamefont {P.}~\bibnamefont {Bøggild}}, \bibinfo {author}
  {\bibfnamefont {R.}~\bibnamefont {Malureanu}}, \ and\ \bibinfo {author}
  {\bibfnamefont {A.}~\bibnamefont {Lavrinenko}},\ }\href@noop {} {\bibfield
  {journal} {\bibinfo  {journal} {Journal of The Electrochemical Society}\
  }\textbf {\bibinfo {volume} {162}},\ \bibinfo {pages} {D213} (\bibinfo {year}
  {2015})}\BibitemShut {NoStop}%
\bibitem [{\citenamefont {Rockstuhl}\ and\ \citenamefont
  {Scharf}(2013)}]{Amorphous}%
  \BibitemOpen
  \bibfield  {author} {\bibinfo {author} {\bibfnamefont {C.}~\bibnamefont
  {Rockstuhl}}\ and\ \bibinfo {author} {\bibfnamefont {T.}~\bibnamefont
  {Scharf}},\ }\href@noop {} {\emph {\bibinfo {title} {Amorphous
  Nanophotonics}}}\ (\bibinfo  {publisher} {Springer Science \& Business
  Media},\ \bibinfo {year} {2013})\BibitemShut {NoStop}%
\bibitem [{\citenamefont {Sihvola}\ and\ \citenamefont
  {of~Electrical~Engineers}(1999)}]{Sihvola1999}%
  \BibitemOpen
  \bibfield  {author} {\bibinfo {author} {\bibfnamefont {A.}~\bibnamefont
  {Sihvola}}\ and\ \bibinfo {author} {\bibfnamefont {I.}~\bibnamefont
  {of~Electrical~Engineers}},\ }\href
  {https://books.google.by/books?id=uIHSNwxBxjgC} {\emph {\bibinfo {title}
  {Electromagnetic Mixing Formulas and Applications}}},\ Electromagnetics and
  Radar Series\ (\bibinfo  {publisher} {Institution of Electrical Engineers},\
  \bibinfo {year} {1999})\BibitemShut {NoStop}%
\bibitem [{\citenamefont {Nielsen}\ \emph {et~al.}(2010)\citenamefont
  {Nielsen}, \citenamefont {Thoreson}, \citenamefont {Chen}, \citenamefont
  {Kristensen}, \citenamefont {Hvam}, \citenamefont {Shalaev},\ and\
  \citenamefont {Boltasseva}}]{Nielsen2010}%
  \BibitemOpen
  \bibfield  {author} {\bibinfo {author} {\bibfnamefont {R.~B.}\ \bibnamefont
  {Nielsen}}, \bibinfo {author} {\bibfnamefont {M.~D.}\ \bibnamefont
  {Thoreson}}, \bibinfo {author} {\bibfnamefont {W.}~\bibnamefont {Chen}},
  \bibinfo {author} {\bibfnamefont {A.}~\bibnamefont {Kristensen}}, \bibinfo
  {author} {\bibfnamefont {J.~M.}\ \bibnamefont {Hvam}}, \bibinfo {author}
  {\bibfnamefont {V.~M.}\ \bibnamefont {Shalaev}}, \ and\ \bibinfo {author}
  {\bibfnamefont {A.}~\bibnamefont {Boltasseva}},\ }\href@noop {} {\bibfield
  {journal} {\bibinfo  {journal} {Applied Physics B}\ }\textbf {\bibinfo
  {volume} {100}},\ \bibinfo {pages} {93} (\bibinfo {year} {2010})}\BibitemShut
  {NoStop}%
\bibitem [{\citenamefont {Zhu}\ and\ \citenamefont {Zhao}(2009)}]{Zhu2009}%
  \BibitemOpen
  \bibfield  {author} {\bibinfo {author} {\bibfnamefont {W.}~\bibnamefont
  {Zhu}}\ and\ \bibinfo {author} {\bibfnamefont {X.}~\bibnamefont {Zhao}},\
  }\href {\doibase 10.1364/JOSAB.26.002382} {\bibfield  {journal} {\bibinfo
  {journal} {J. Opt. Soc. Am. B}\ }\textbf {\bibinfo {volume} {26}},\ \bibinfo
  {pages} {2382} (\bibinfo {year} {2009})}\BibitemShut {NoStop}%
\bibitem [{\citenamefont {Andryieuski}\ \emph {et~al.}(2014)\citenamefont
  {Andryieuski}, \citenamefont {Zhukovsky},\ and\ \citenamefont
  {Lavrinenko}}]{Andryieuski2014}%
  \BibitemOpen
  \bibfield  {author} {\bibinfo {author} {\bibfnamefont {A.}~\bibnamefont
  {Andryieuski}}, \bibinfo {author} {\bibfnamefont {S.~V.}\ \bibnamefont
  {Zhukovsky}}, \ and\ \bibinfo {author} {\bibfnamefont {A.~V.}\ \bibnamefont
  {Lavrinenko}},\ }\href@noop {} {\bibfield  {journal} {\bibinfo  {journal}
  {Optics express}\ }\textbf {\bibinfo {volume} {22}},\ \bibinfo {pages}
  {14975} (\bibinfo {year} {2014})}\BibitemShut {NoStop}%
\bibitem [{\citenamefont {van Lare}\ and\ \citenamefont
  {Polman}(2015)}]{Polman2015}%
  \BibitemOpen
  \bibfield  {author} {\bibinfo {author} {\bibfnamefont {M.-C.}\ \bibnamefont
  {van Lare}}\ and\ \bibinfo {author} {\bibfnamefont {A.}~\bibnamefont
  {Polman}},\ }\href@noop {} {\bibfield  {journal} {\bibinfo  {journal} {ACS
  Photonics}\ }\textbf {\bibinfo {volume} {2}},\ \bibinfo {pages} {822}
  (\bibinfo {year} {2015})}\BibitemShut {NoStop}%
\bibitem [{\citenamefont {Helgert}\ \emph {et~al.}(2009)\citenamefont
  {Helgert}, \citenamefont {Rockstuhl}, \citenamefont {Etrich}, \citenamefont
  {Menzel}, \citenamefont {Kley}, \citenamefont {T\"unnermann}, \citenamefont
  {Lederer},\ and\ \citenamefont {Pertsch}}]{Helgert2009}%
  \BibitemOpen
  \bibfield  {author} {\bibinfo {author} {\bibfnamefont {C.}~\bibnamefont
  {Helgert}}, \bibinfo {author} {\bibfnamefont {C.}~\bibnamefont {Rockstuhl}},
  \bibinfo {author} {\bibfnamefont {C.}~\bibnamefont {Etrich}}, \bibinfo
  {author} {\bibfnamefont {C.}~\bibnamefont {Menzel}}, \bibinfo {author}
  {\bibfnamefont {E.-B.}\ \bibnamefont {Kley}}, \bibinfo {author}
  {\bibfnamefont {A.}~\bibnamefont {T\"unnermann}}, \bibinfo {author}
  {\bibfnamefont {F.}~\bibnamefont {Lederer}}, \ and\ \bibinfo {author}
  {\bibfnamefont {T.}~\bibnamefont {Pertsch}},\ }\href {\doibase
  10.1103/PhysRevB.79.233107} {\bibfield  {journal} {\bibinfo  {journal} {Phys.
  Rev. B}\ }\textbf {\bibinfo {volume} {79}},\ \bibinfo {pages} {233107}
  (\bibinfo {year} {2009})}\BibitemShut {NoStop}%
\bibitem [{\citenamefont {M\"{u}hlig}\ \emph {et~al.}(2011)\citenamefont
  {M\"{u}hlig}, \citenamefont {Rockstuhl}, \citenamefont {Yannopapas},
  \citenamefont {B\"{u}rgi}, \citenamefont {Shalkevich},\ and\ \citenamefont
  {Lederer}}]{Muhlig2011}%
  \BibitemOpen
  \bibfield  {author} {\bibinfo {author} {\bibfnamefont {S.}~\bibnamefont
  {M\"{u}hlig}}, \bibinfo {author} {\bibfnamefont {C.}~\bibnamefont
  {Rockstuhl}}, \bibinfo {author} {\bibfnamefont {V.}~\bibnamefont
  {Yannopapas}}, \bibinfo {author} {\bibfnamefont {T.}~\bibnamefont
  {B\"{u}rgi}}, \bibinfo {author} {\bibfnamefont {N.}~\bibnamefont
  {Shalkevich}}, \ and\ \bibinfo {author} {\bibfnamefont {F.}~\bibnamefont
  {Lederer}},\ }\href {\doibase 10.1364/OE.19.009607} {\bibfield  {journal}
  {\bibinfo  {journal} {Opt. Express}\ }\textbf {\bibinfo {volume} {19}},\
  \bibinfo {pages} {9607} (\bibinfo {year} {2011})}\BibitemShut {NoStop}%
\bibitem [{\citenamefont {Albooyeh}\ \emph {et~al.}(2012)\citenamefont
  {Albooyeh}, \citenamefont {Morits},\ and\ \citenamefont
  {Tretyakov}}]{Albooyeh2012}%
  \BibitemOpen
  \bibfield  {author} {\bibinfo {author} {\bibfnamefont {M.}~\bibnamefont
  {Albooyeh}}, \bibinfo {author} {\bibfnamefont {D.}~\bibnamefont {Morits}}, \
  and\ \bibinfo {author} {\bibfnamefont {S.~A.}\ \bibnamefont {Tretyakov}},\
  }\href {\doibase 10.1103/PhysRevB.85.205110} {\bibfield  {journal} {\bibinfo
  {journal} {Phys. Rev. B}\ }\textbf {\bibinfo {volume} {85}},\ \bibinfo
  {pages} {205110} (\bibinfo {year} {2012})}\BibitemShut {NoStop}%
\bibitem [{\citenamefont {Albooyeh}\ \emph {et~al.}(2014)\citenamefont
  {Albooyeh}, \citenamefont {Kruk}, \citenamefont {Menzel}, \citenamefont
  {Helgert}, \citenamefont {Kroll}, \citenamefont {Krysinski}, \citenamefont
  {Decker}, \citenamefont {Neshev}, \citenamefont {Pertsch}, \citenamefont
  {Etrich}, \citenamefont {Rockstuhl}, \citenamefont {Tretyakov}, \citenamefont
  {Simovski},\ and\ \citenamefont {Kivshar}}]{Albooyeh2014}%
  \BibitemOpen
  \bibfield  {author} {\bibinfo {author} {\bibfnamefont {M.}~\bibnamefont
  {Albooyeh}}, \bibinfo {author} {\bibfnamefont {S.}~\bibnamefont {Kruk}},
  \bibinfo {author} {\bibfnamefont {C.}~\bibnamefont {Menzel}}, \bibinfo
  {author} {\bibfnamefont {C.}~\bibnamefont {Helgert}}, \bibinfo {author}
  {\bibfnamefont {M.}~\bibnamefont {Kroll}}, \bibinfo {author} {\bibfnamefont
  {A.}~\bibnamefont {Krysinski}}, \bibinfo {author} {\bibfnamefont
  {M.}~\bibnamefont {Decker}}, \bibinfo {author} {\bibfnamefont {D.~N.}\
  \bibnamefont {Neshev}}, \bibinfo {author} {\bibfnamefont {T.}~\bibnamefont
  {Pertsch}}, \bibinfo {author} {\bibfnamefont {C.}~\bibnamefont {Etrich}},
  \bibinfo {author} {\bibfnamefont {C.}~\bibnamefont {Rockstuhl}}, \bibinfo
  {author} {\bibfnamefont {S.~a.}\ \bibnamefont {Tretyakov}}, \bibinfo {author}
  {\bibfnamefont {C.~R.}\ \bibnamefont {Simovski}}, \ and\ \bibinfo {author}
  {\bibfnamefont {Y.~S.}\ \bibnamefont {Kivshar}},\ }\href {\doibase
  10.1038/srep04484} {\bibfield  {journal} {\bibinfo  {journal} {Scientific
  reports}\ }\textbf {\bibinfo {volume} {4}},\ \bibinfo {pages} {4484}
  (\bibinfo {year} {2014})}\BibitemShut {NoStop}%
\bibitem [{\citenamefont {Tretyakov}(2003)}]{Tretyakov2003}%
  \BibitemOpen
  \bibfield  {author} {\bibinfo {author} {\bibfnamefont {S.}~\bibnamefont
  {Tretyakov}},\ }\href {https://books.google.by/books?id=MZ3tpGtadhcC} {\emph
  {\bibinfo {title} {Analytical Modeling in Applied Electromagnetics}}},\
  Artech House electromagnetic analysis series\ (\bibinfo  {publisher} {Artech
  House},\ \bibinfo {year} {2003})\BibitemShut {NoStop}%
\end{thebibliography}%

\appendix

\section{Supercell approach}

In the supercell approach we replace the random array with a quasi-random array, namely, with a periodic array of square $N\times N$-cells supercells, and within each supercell the particles are chosen randomly [see Fig. \ref{Fig03} (a)]. The particles at identical positions number $(m,n)$ within different supercells are identical and under normally incident plane wave illumination they have identical dipole moments.
\begin{eqnarray} \label{Eq:periodicity}
\alpha_{m,n}=\alpha_{m+N,n}=\alpha_{m,n+N}, \\
p_{m,n}=p_{m+N,n}=p_{m,n+N}.
\end{eqnarray}
For simplicity of numeration we can always consider the particle $(m,n)$ as the center of a  virtual moving frame (red dashed rectangle in Fig.~\ref{Fig03} (a)) of the size $N\times N$. Therefore we can temporary call $p_{m,n}\equiv \tilde{p}_{0,0}$ and $\alpha_{m,n}\equiv \tilde{\alpha}_{0,0}$. Moreover, we can require that $N$ is an odd integer $N=2N_0+1$, where $N_0$ is an integer number.

We consider normally incident plane waves of the amplitude $E_0$. The dipole moment of the considered particle can be written through the infinite sum
\begin{equation} \label{Eq:infinite_set}
\tilde{p}_{0,0}=\tilde{\alpha}_{0,0} \left(E_0+\sum_{m'=-\infty}^{+\infty}\sum_{n'=-\infty}^{+\infty}b_{m',n'}\tilde{p}_{m',n'}\right),
\end{equation}
\noindent where the dipole coefficients are
\begin{equation}
b_{0,0}=0,
\end{equation}
\begin{multline}
b_{m',n'}=\frac{1}{4\pi} \frac{1}{\varepsilon_0 a^3} \exp(ikaR') \frac{1}{R'^3} \times \\ 
\times \left[ k^2 a^2 m'^2 + (2n'^2-m'^2)(\frac{1}{R'^2} - i\frac{ka}{R'})\right],
\end{multline}
\noindent and $R'=\sqrt{m'^2+n'^2}$ and $k=2\pi/\lambda$. One may notice that 
\begin{equation}
b_{m',n'}=b_{-m',n'}=b_{m',-n'}=b_{-m',-n'}.
\end{equation}

Using the periodicity of the supercell and introducing the coefficients
\begin{equation}
\beta_{m',n'}=\sum_{M'=-\infty}^{+\infty} \sum_{N'=-\infty}^{+\infty} b_{m'+NM',n'+NN'},
\end{equation}
\noindent we reformulate the Eq. (\ref{Eq:infinite_set}) to a finite sum and obtain the equation for the dipole moment
\begin{equation}
\frac{1}{\tilde{\alpha}_{0,0}} \tilde{p}_{0,0} - \sum_{m'=-N_0}^{+N_0}\sum_{n'=-N_0}^{+N_0}\beta_{m',n'}\tilde{p}_{m',n'}=E_0.
\end{equation}
Such equation can be written for each particle within the supercell. If we renumerate the initial particle $(m,n)$ within the square supercell so that its number $l=(m-1)N+n$ and using the periodicity Eq. (\ref{Eq:periodicity}) and the fact that $\beta_{m',n'}=\beta_{-m',n'}$ and applying the correct index shift, we get a system of $N\times N$ set of linear equations for determination of the dipole moments of each particle within the system:
\begin{equation}
\label{Eq:MatrixEq}
(\boldsymbol{A}-\boldsymbol{B})\left[\boldsymbol{p}\right]=E_0\left[\boldsymbol{1}\right],
\end{equation}
\noindent where
\[
\boldsymbol{A}=
\begin{bmatrix} 
\frac{1}{\alpha_1} & 0 & 0 & \cdots & 0 \\ 
0 & \frac{1}{\alpha_2} & 0 & \cdots & 0 \\
0 & 0 & \frac{1}{\alpha_3} & \cdots & 0 \\
\vdots & \vdots & \vdots & \ddots & \vdots\\ 
0 & 0 & 0 & \cdots & \frac{1}{\alpha_{NN}}
\end{bmatrix},
\]

\[
 \boldsymbol{B}= 
 \begin{bmatrix} 
\beta_{0,0} & \beta_{0,1} & \beta_{0,2} & \cdots & \beta_{0,1} \\ 
\beta_{0,1} & \beta_{0,0} & \beta_{0,1} & \cdots & \beta_{0,2} \\
\beta_{0,2} & \beta_{0,1} & \beta_{0,0} & \cdots & \beta_{0,3} \\
\vdots & \vdots & \vdots & \ddots & \vdots\\ 
\beta_{0,1} & \beta_{0,2} & \beta_{0,3} & \cdots & \beta_{0,0}
\end{bmatrix},
\]

\[
\left[\boldsymbol{p}\right]=
\begin{bmatrix} 
p_{1} \\ 
p_{2} \\
p_{3} \\
\vdots \\ 
p_{NN}
\end{bmatrix},
\] 
and
\[
\left[\boldsymbol{1}\right]=
\begin{bmatrix} 
1 \\ 
1 \\
1 \\
\vdots \\ 
1
\end{bmatrix}.
\]

Practically, when calculating coefficients $\beta_{m',n'}$ we cannot sum an infinite number of addends and have to limit summation at a certain number $N_{\mathrm{max}}$
\begin{equation}
\beta_{m',n'}=\sum_{M'=-N_{\mathrm{max}}}^{+N_{\mathrm{max}}} \sum_{N'=-N_{\mathrm{max}}}^{+N_{\mathrm{max}}} b_{m'+NM',n'+NN'}.
\end{equation}
In other words, instead of an infinite number of supercells we consider an array of $(2N_\mathrm{max}+1)\times(2N_\mathrm{max}+1)$ supercells and free space outside. This abrupt array termination causes a problem, since the series for $\beta_{m',n'}$ is poorly converging due to oscillating complex exponential in dipole coefficient $b_{m',n'}$. We used $N_\mathrm{max}$ up to 500 and have not observed satisfactory convergence. The finite array interaction constant $\beta^*=\sum_{m'=-N_0}^{+N_0}\sum_{n'=-N_0}^{+N_0}\beta_{m',n'}$ gives a large relative error of 5\% with respect to  interaction constant $\beta$ for an infinite array \ref{eq:beta} and such error in $\beta$ may result in the loss factor error larger than the influence of diffuse scattering. This problem was solved by adjusting $\beta_{m',n'}$ coefficients, namely
\begin{equation}
\beta_{m',n'}=\sum_{M'=-N_{\mathrm{max}}}^{+N_{\mathrm{max}}} \sum_{N'=-N_{\mathrm{max}}}^{+N_{\mathrm{max}}} b_{m'+NM',n'+NN'}+(\beta-\beta^*)/N^2.
\end{equation}
Obviously, coefficients $\beta_{m',n'}$ defined like this strictly give the correct value of the interaction constant for an infinite array
\begin{equation}
\sum_{M'=-N_0}^{+N_0} \sum_{N'=-N_0}^{+N_0} \beta_{m',n'}=\beta.
\end{equation}
It is easy to show that physically this "adjustment" of $\beta_{m',n'}$ is equivalent to assuming that all particles outside the $(2N_{\mathrm{max}}+1)×(2N_{\mathrm{max}}+1)$ supercells array have identical dipole moments equal to the mean dipole moment within a supercell 
\begin{equation}
\langle p \rangle = \frac{1}{N^2} \sum_{m=-N_0}^{+N_0}\sum_{n=-N_0}^{+N_0}p_{m,n}.
\end{equation}
This is obviously a more physically sound approximation to a completely infinite supercells array than the case of an abrupt array termination, leading to parasitic scattering from the array edges.

Solving system Eq. (\ref{Eq:MatrixEq}) we find all the dipole moments within the supercell and are able to calculate the transmitted, reflected and diffusely scattered powers.

\section{Calculation of the diffuse scattering losses}

Assume that we know the dipole moment $p_{m,n}$ of each particle number $(m,n)$ within the supercell. In the simulations we selected supercells larger than the wavelength $\lambda$ in order to observe the effects of diffuse scattering and to have a sufficient number of random particles within each supercell. Thus, the array of the supercells is a kind of diffraction grating. We can calculate explicitly the transmitted, reflected and scattered fields by this diffraction grating.

First, we consider the dipole $p=p_0exp(-i\omega t)$ in the center of the coordinate system oriented as shown in Fig. \ref{Fig03} (b). We assume the wave incidence along the $z-$axis. The Cartesian coordinates $(x,y,z)$ of a point in space are related to its spherical coordinates as $(r,\theta,\phi)$ as $x=r\sin\theta \sin\phi, y=r\cos\theta, z=r\sin\theta\cos\phi$.

The electromagnetic field of the dipole in the far zone in the spherical coordinate system is
\begin{eqnarray}
H_r=H_\theta=E_\phi=0, \\
H_\phi=-\frac{\omega p_0 k \sin\theta}{4\pi r} e^{ikr},\\
E_r=-\frac{i\omega p_0 \eta k \cos\theta}{2\pi r} e^{ikr},\\
E_\theta=-\frac{\omega p_0 \eta k \sin\theta}{4\pi r} e^{ikr}=\eta H_\phi,
\end{eqnarray}
\noindent where $\eta = 120\pi [\Omega]$ is the free-space impedance and $k=\omega/c=2\pi/\lambda$ is the wavenumber.

The amplitude of the average Poynting vector directed outside the sphere with the dipole in the center is
\begin{equation}
S_r=\frac{1}{2}  E_\theta H_\phi^*=\eta E_\theta E_\theta^*.
\end{equation}
In order to find the total field scattered by the supercell array we need to sum up only the components $E_{\theta}$ of all the dipoles.

We can select the coordinate systems in such a way that the coordinates of a particle number $(m,n)$ within the supercell are $x_m=ma, y_n = na$. Due to randomness of the array all the dipole moments differ in the amplitude and phase. Moreover, for a specific direction of the scattered plane wave propagation with the wavevector $(k_x, k_y, k_z)$, where $k_z=\sqrt{k^2-k_x^2-k_y^2}$ there is a geometrical phase shift $\Delta\chi_{m,n}$ for the point $(x_m,y_n)$ with respect to the coordinate system origin $(0,0)$
\begin{equation}
\Delta\chi_{m,n}=k_x am+k_y an.
\end{equation}
The total field created by the supercell is
\begin{multline}
E_\theta^\mathrm{supercell}(r,k_x,k_y)=-\frac{\omega p_0 \eta k \sin\theta}{4\pi r} e^{ikr} \times \\
\times\sum_{m=1}^{N} \sum_{n=1}^{N}p_{m,n}e^{i\Delta\chi_{m,n}},
\end{multline}

\noindent where $\sin\theta=\sqrt{1-k_y^2/k^2}$. Let's call $\Sigma_p(k_x,k_y)\equiv\sum_{m=1}^{N} \sum_{n=1}^{N}p_{m,n}e^{i\Delta\chi_{m,n}}$, then

\begin{equation}
E_\theta^\mathrm{supercell}(r,k_x,k_y)=-\frac{\omega p_0 \eta k \sin\theta}{4\pi r} e^{ikr} \Sigma_p(k_x,k_y).
\end{equation}

Next, we should sum up the fields created by various supercells, taking into account the phase advance between them. The phase advance for the supercell number $(m^*,n^*)$ is $\Delta\chi_{m^*,n^*}=(k_x m^*+k_y n^*)Na$. The total field for $M^*\times N^*$ supercells is
\begin{multline}
E_\theta^\mathrm{total}(r,k_x,k_y)=\sum_{m^*=1}^{M^*} \sum_{n^*=1}^{N^*}E_\theta^\mathrm{supercell}(r,k_x,k_y)\times\\
\times e^{i(k_x m^*+k_y n^*)Na}.
\end{multline}

Using the following identities
\begin{multline}
\sum_{m^*=1}^{M^*}e^{ik_x m^*Na} = e^{iNak_x(M^*+1)/2}\frac{\sin\frac{Nak_x M^*}{2}}{\sin\frac{Nak_x}{2}},\\
\sum_{n^*=1}^{N^*}e^{ik_y n^*Na} = e^{iNak_y(N^*+1)/2}\frac{\sin\frac{Nak_y N^*}{2}}{\sin\frac{Nak_y}{2}},
\end{multline}
\noindent we find the Poynting vector in direction $(k_x,k_y)$:
\begin{multline}
S_r (r,k_x,k_y)=\frac{1}{2\eta} |E_{\theta}^\mathrm{supercell}(r,k_x,k_y)|^2 \times \\
\times \left|\frac{\sin\frac{Nak_x M^*}{2}}{\sin\frac{Nak_x}{2}}\right|^2 \left|\frac{\sin\frac{Nak_y N^*}{2}}{\sin\frac{Nak_y}{2}}\right|^2.
\end{multline}

After some tedious transformations we find that the power emitted from a unit area in direction $(k_x,k_y)$ is
\begin{multline}
dS = \frac{\eta\omega^4}{32c^2} |\Sigma_p(k_x,k_y)|^2 \frac{1}{N^2a^2} \frac{1-k_y^2/k^2}{k^2\sqrt{1-k_x^2/k^2-k_y^2/k^2}}dk_xdk_y \times \\
\times \frac{1}{\pi M^*} \left|\frac{\sin\frac{Nak_x M^*}{2}}{\sin\frac{Nak_x}{2}}\right|^2 \frac{1}{\pi N^*} \left|\frac{\sin\frac{Nak_y N^*}{2}}{\sin\frac{Nak_y}{2}}\right|^2.
\end{multline}
However, for an infinite array $M^*\to\infty, N^*\to\infty$ and this expression can be simplified since
$$
\lim_{M^*\to\infty} \frac{1}{\pi M^*} \left|\frac{\sin\frac{Nak_x M^*}{2}}{\sin\frac{Nak_x}{2}}\right|^2 = \frac{2}{Na}\sum_{\tilde{m}}\delta(k_x-\frac{2\pi\tilde{m}}{Na}),
$$

$$
\lim_{N^*\to\infty} \frac{1}{\pi N^*} \left|\frac{\sin\frac{Nak_y N^*}{2}}{\sin\frac{Nak_y}{2}}\right|^2 = \frac{2}{Na}\sum_{\tilde{n}}\delta(k_y-\frac{2\pi\tilde{n}}{Na}),
$$
\noindent where $\delta(x)$ is the Dirac delta-function and summation is done for the propagating waves only $k_{x,\tilde{m}}^2+k_{y,\tilde{n}}^2\leq k^2$ and thus $\tilde{m}^2+\tilde{n}^2\leq (\frac{Na}{\lambda})^2$. The delta-functions in the expression represent the properties of the diffraction grating to scatter light only in specific directions. Finally, we get
\begin{multline}
dS = \frac{\eta\omega^2}{8} |\Sigma_p(k_x,k_y)|^2 \frac{1}{N^4a^4} \frac{1-k_y^2/k^2}{k^2\sqrt{1-k_x^2/k^2-k_y^2/k^2}} \times \\
\times \sum_{\tilde{m}}\delta(k_x-\frac{2\pi\tilde{m}}{Na}) \sum_{\tilde{n}}\delta(k_y-\frac{2\pi\tilde{n}}{Na}) dk_xdk_y.
\end{multline}
In order to calculate the normally reflected power one has to integrate the expression 
\begin{multline}
S_\mathrm{refl}=\int_{-0}^{+0}dk_x\int_{-0}^{+0}dk_y \frac{\eta\omega^2}{8} |\Sigma_p(k_x,k_y)|^2 \frac{1}{N^4a^4} \times \\
\times \frac{1-k_y^2/k^2}{k^2\sqrt{1-k_x^2/k^2-k_y^2/k^2}}\delta(k_x)\delta(k_y) =\\
= \frac{\eta\omega^2}{8} |\Sigma_p(0,0)|^2 \frac{1}{N^4a^4}. 
\end{multline}
In order to calculate the reflectance one has to divide the reflected power flux by the incident power flux $S_\mathrm{inc}=E_0^2/2\eta$:
\begin{multline}
R=S_\mathrm{refl}/S_\mathrm{inc}=\frac{\eta^2\omega^2}{4} |\Sigma_p(0,0)|^2 \frac{1}{N^4a^4E_0^2}=\\
=\frac{k^2a^2}{4} \frac{|\langle p \rangle|^2}{\varepsilon_0^2a^6E_0^2},
\end{multline}

\noindent where $\langle p \rangle = \Sigma_p(0,0)/N^2$ is the mean dipole moment of a subcell of the size $a\times a$. This result is in  agreement with the reflectance for a regular array of regular dipoles with the dipole moment $p$ created by the incident wave $E_0$, namely, $R=\frac{k^2a^2}{4} \frac{|p|^2}{\varepsilon_0^2a^6E_0^2}$. The amplitude reflection is
\begin{equation}
r=\frac{ika}{2}\frac{\langle p \rangle}{\varepsilon_0a^3E_0}.
\end{equation}
The amplitude and power transmission coefficients are
\begin{eqnarray}
t=1+r=1+\frac{ika}{2}\frac{\langle p \rangle}{\varepsilon_0a^3E_0},\\
T=|t|^2=\frac{k^2a^2}{4}|\frac{2}{ika}+\frac{\langle p \rangle}{\varepsilon_0a^3E_0}|^2.
\end{eqnarray}

In order to calculate the scattered power one has to integrate over  $k_x, k_y$ within a circle $k_x^2+k_y^2\leq k^2$, excluding the point $(k_x=0,k_y=0)$:
\begin{multline}
S_\mathrm{diffuse}=2\times[-S_\mathrm{refl}+{\int\int}_{k_x^2+k_y^2\leq k^2} dk_xdk_y \frac{\eta\omega^2}{8} |\Sigma_p(k_x,k_y)|^2 \times \\ 
\times\frac{1}{N^4a^4} \frac{1-k_y^2/k^2}{k^2\sqrt{1-k_x^2/k^2-k_y^2/k^2}} \times \\
\times \sum_{\tilde{m}}\delta(k_x-\frac{2\pi\tilde{m}}{Na}) \sum_{\tilde{n}}\delta(k_y-\frac{2\pi\tilde{n}}{Na})]. 
\end{multline}
The factor $2$ comes from the fact that diffuse scattering happens in both $+k_z$ and $-k_z$ directions. The loss factor attributed to the diffuse scattering is then 
\begin{multline}
A_\mathrm{diffuse}=S_\mathrm{diffuse}/S_\mathrm{inc}=2\times [-R+\frac{k^2a^2}{4} \times \\ 
\times \sum_{\tilde{m}^2+\tilde{n}^2\leq (Na/\lambda)^2}|\frac{\Sigma_p(k_{x\tilde{m}},k_{y\tilde{n}})}{N^2\varepsilon_0a^3E_0}|\frac{1-k_{y\tilde{n}}^2/k^2}{\sqrt{1-k_{x\tilde{m}}^2/k^2-k_{y\tilde{n}}^2/k^2}}],
\end{multline}

\noindent where $k_{x\tilde{m}}=2\pi\tilde{m}/Na$ and  $k_{y\tilde{n}}=2\pi\tilde{n}/Na$.

\end{document}